\documentclass{PoS}
\usepackage{epsfig}

\PoS{PoS(BDMH2004)002}

\title{Searching for the Missing Baryons in the Warm-hot Intergalactic Medium}

\ShortTitle{Searching for the Missing Baryons in the WHIM}

\author{\speaker{Philipp Richter}\thanks{DFG Emmy-Noether Fellow}\\
        IAEF Bonn, Germany\\
        E-mail: \email{prichter@astro.uni-bonn.de}}

\author{Blair D. Savage\\
        Department of Astronomy, University of Wisconsin-Madison, USA}

\author{Todd M. Tripp\\
        Department of Astronomy, University of Massachusetts, USA}

\author{Kenneth R. Sembach\\
        Space Telescope Science Institute, USA}

\abstract{
We discuss physical properties and the baryonic content of 
the Warm-hot Intergalactic Medium (WHIM) at low redshifts.
Cosmological simulations predict that the WHIM contains a large fraction
of the baryons at $z=0$ in the form of highly-ionized gas
at temperatures between $10^5$ and $10^7$ K. Using
high-resolution ultraviolet spectra obtained with the
{\it Space Telescope Imaging Spectrograph} (STIS) and the
{\it Far Ultraviolet Spectroscopic Explorer} (FUSE) we have
studied the WHIM at low redshifts by searching for intervening
O\,{\sc vi} and thermally broadened Lyman $\alpha$ (BL) absorption toward 
a number of quasars and active galactic nuclei (AGNs). Our 
measurements imply cosmological mass densities of 
$\Omega_b$(O\,{\sc vi})$\approx0.0027$ $h_{75}^{-1}$ and
$\Omega_{b}$(BL)$\approx0.0058\,h_{75}^{-1}$.
Our results suggest that the WHIM at low $z$ contains more baryonic
mass than stars and gas in galaxies.
}
\FullConference{Baryons in Dark Matter Halos\\
                5-9 October 2004\\
                Novigrad, Croatia}

\begin{document}

\section{Introduction}

Weakly and highly ionized intergalactic gas
most likely makes up for most of the baryonic matter in the local
Universe. While the diffuse photoionized intergalactic
medium (IGM) that gives rise to the Lyman $\alpha$
forest is expected to account for $\sim 30$ percent of the
baryons today (Penton, Stocke, \& Shull 2004 [1]), 
the highly-ionized {\it Warm-Hot Intergalactic Medium} (WHIM)
at temperatures $T\sim 10^5-10^7$ K possibly contributes
at a comparable level to the cosmological mass density of
the baryons at $z=0$ (e.g., Dav\a'e et al.\,2001\,[2]).
Gas and stars in and around galaxies and galaxy
clusters make up the rest (Fukugita 2003\,[3]).
The total cosmological mass density of the baryons, $\Omega_b$, has
been restricted to a value of $\Omega_b\approx0.04$,
e.g., by measuring the deuterium-to-hydrogen ratio in high-redshift
absorption line systems (Burles \& Tytler 1998\,[4]) and by analyzing
the small-scale anisotropy of the cosmic-microwave background (Spergel
et al.\,2003 [5]). 
However, the contribution 
from the WHIM to $\Omega_b$ has been estimated 
mainly from numerical simulations of the structure
formation in the Universe rather than from observational results
(e.g., Cen \& Ostriker 1999\,[6]; Dav\a'e et al.\,2001\,[2]). 
So far, only a few percent of the
baryonic matter residing in the WHIM at $z=0$ actually has been
detected.
To test the cosmological simulations and to
learn about the distribution of the baryons in the local
Universe it is of crucial importance to look for possibilities
to pinpoint the baryon budget in the WHIM by direct observations.

\section{O\,{\sc vi} Absorbers}

Five-times ionized
oxygen (O\,{\sc vi}) currently is the most important high ion
to trace the WHIM at temperatures of
$T\sim 3\times 10^5$ K
in the FUV regime. Oxygen is a
relatively abundant element and the two available O\,{\sc vi}
transitions (located at $1031.9$ and $1037.6$ \AA) have
large oscillator strengths.
A number of detections
of intervening WHIM O\,{\sc vi} absorbers at $z<0.5$
have been reported in the literature
(see Sembach et al.\,2004\,[7] and references therein)
based on observations with HST/STIS and FUSE.
These measurements imply a number density of O\,{\sc vi}
absorbers per unit redshift of $dN_{\rm OVI}/dz \approx 14$
for equivalent widths $W_{\lambda}\geq50$ m\AA, as derived
from the analysis of six lines of sight
(Sembach et al.\,2004\,[7]).
Assuming that 20 percent or less of the oxygen is present
in the form of O\,{\sc vi} ($f_{\rm O\,VI} \leq 0.2$)
and further assuming a mean oxygen abundance of $0.1$ solar,
the measured number density
corresponds to a cosmologial mass density
of $\Omega_b$(O\,{\sc vi})$\approx0.0027$ $h_{75}^{-1}$.
For the interpretation of this value it has to be noted
that O\,{\sc vi} absorption mainly traces gas with
temperatures around $3 \times 10^5$ K, but not the
million-degree gas phase which probably contains
the majority of the baryons in the WHIM.
Very recently, Savage et al.\,(2004\,[8]) have reported the
detection of Ne\,{\sc viii} in an absorption system
at $z\approx 0.2$ in the direction of the quasar HE\,0226$-$4110.
Ne\,{\sc viii} traces gas at $T\sim 7\times10^5$ K and thus
is possibly suited to complement the O\,{\sc vi} measurements
of the WHIM in a higher temperature regime. However, as the
cosmic abundance of neon is relatively low, it is not expected
that Ne\,{\sc viii} is a particularly sensitive tracer of
the WHIM. This is supported by the non-detections of intervening Ne\,{\sc viii}
in other high S/N STIS data (Richter et al.\,2004\,[9]).

\begin{figure}[t!]
\epsfig{file=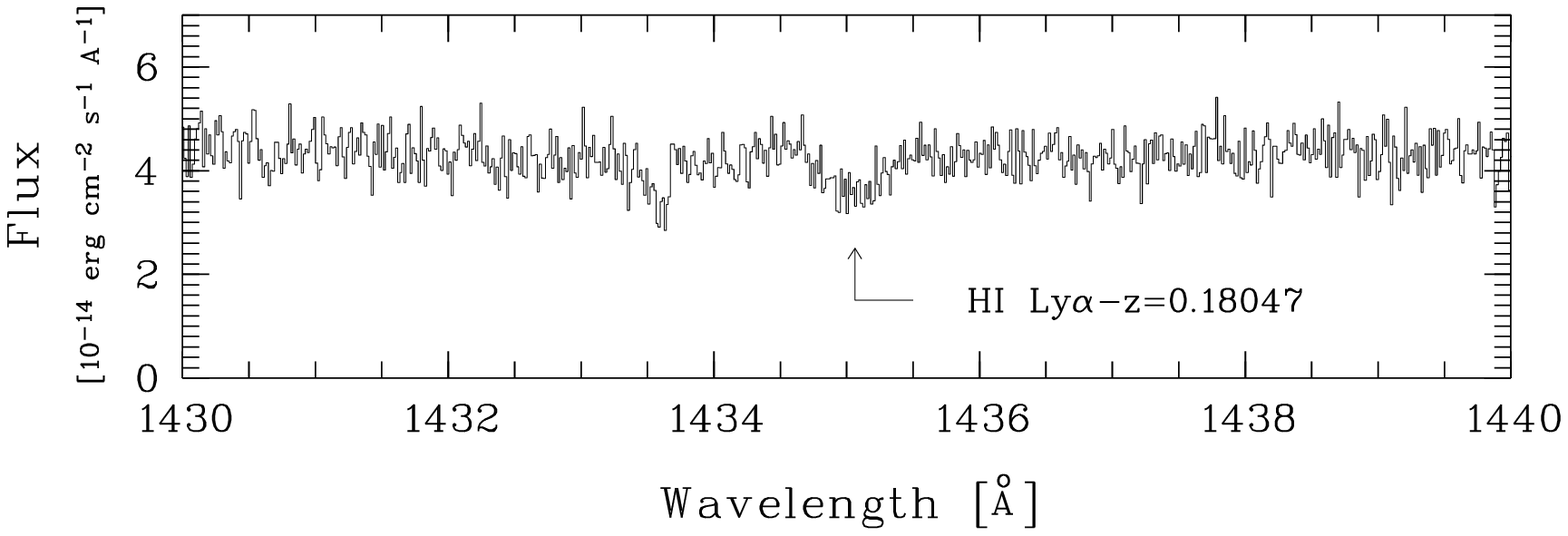, width=1.0\textwidth}
\caption{
Broad Ly\,$\alpha$ absorber in the STIS spectrum of H\,1821+643 at
$z=0.18047$. This absorber has a $b$ value of $56$ km\,s$^{-1}$ and 
an H\,{\sc i} column density of log $N=13.21$. For pure thermal broadening
and assuming CIE, these numbers imply a gas temperature of $\sim 2 \times
10^5$ K and a total hydrogen column density (neutral + ionized) of
log $N=18.61$.
}
\end{figure}

\newpage
\section{Broad Lyman $\alpha$ Absorbers}

Next to high-ion absorption from oxygen and other metals,
recent observations with STIS (Richter et al.\,2004\,[9]; Sembach
et al.\,2004\,[7]) suggest that WHIM filaments can be detected
in Ly\,$\alpha$ absorption of {\it neutral} hydrogen.
Although the vast majority of the hydrogen in the WHIM
is ionized, a tiny fraction
($<10^{-5}$, typically) of neutral
hydrogen should be present if the gas is
in collisional ionization equilibrium (CIE; see
Sutherland \& Dopita 1993\,[10]). Depending on the total
gas column density of a WHIM absorber and its
temperature, weak H\,{\sc i} Ly\,$\alpha$ absorption
at column densites $12.5\leq$ log $N$(H\,{\sc i})$\leq 14.0$
may arise from WHIM filaments and could be used to
trace the overlaying ionized hydrogen component.
The Ly\,$\alpha$ absorption from WHIM filaments is
expected to be very broad due to thermal line
broadening, resulting in large Doppler parameters
of $b>40$ km\,s$^{-1}$. 
For pure thermal broadening, the $b$ value of the Ly\,$\alpha$
line is related to the temperature of the gas via
log $T\approx\,{\rm log}\,(60\,b^2)$.
In CIE, the ionized
gas fraction ($f_{\rm H}=$H$_{\rm total}/$H\,{\sc i}) can be approximated via
${\rm log}\,f_{\rm H}(T)\approx -13.9 + 5.4\,{\rm log}\,T - 0.33\,
({\rm log}\,T)^2$ (Richter et al.\,2004\,[9]).
Under the assumption that thermal broadening dominates
the widths of the WHIM Ly\,$\alpha$ absorbers, one then
can derive the total hydrogen mass of an individual
absorber from its $b$ value and H\,{\sc i} column density. 
However, intergalactic  Ly\,$\alpha$ absorbers with 
$b$ values $>40$ km\,s$^{-1}$ are generally difficult
to detect, as they are broad and shallow. High resolution,
high S/N FUV spectra of QSOs with smooth background continua
are required to successfully search for broad Ly\,$\alpha$
absorption in the low-redshift WHIM. 
STIS (functional until 2004) was
the only instrument available that delivered such data,
but due to the instrumental limitations of space-based
observatories, the number of QSO spectra eligible for searching
for WHIM broad Ly\,$\alpha$ absorbers
is very limited.
The cosmological mass density of the broad Ly\,$\alpha$ absorbers
at low $z$, $\Omega_b{\rm (BL)}$, can
be derived from the observed number statistics of broad 
Ly\,$\alpha$ absorbers in high-resolution STIS data.
We can write:
\begin{equation}
\Omega_b{\rm (BL)}=\frac{\mu\,m_{\rm H}\,H_0}
{\rho_{\rm c}\,c}\,\sum_{ij}\,f_{{\rm H},ij}\,N({\rm H\,I})_{ij}\,
\Big / \sum_{j}\Delta X_j,
\end{equation}
with $\mu=1.3$, $m_{\rm H}=1.673 \times 10^{-27}$ kg, $H_0=
75$ km\,s$^{-1}$\,Mpc$^{-1}$, and $\rho_{\rm c}=3H_0\,^2/8 \pi G$.
The index $i$ denotes an individual broad Ly\,$\alpha$ system
along a line of sight $j$. Each measured absorption
system $i$ is characterized
by its neutral hydrogen column density, $N$(H\,{\sc i})$_{ij}$, and
ionization fraction, $f_{{\rm H},ij}$.
Each line of sight $j$ has a
characteristic redshift range, $\Delta X_j$,
available for detecting broad Ly\,$\alpha$
absorption (see, e.g., Sembach et al.\,2004\,[7]).

We recently have measured the number density of 
broad Ly\,$\alpha$ absorbers at low $z$ in the directions
of the quasars PG\,1259+593, PG\,1116+215, H\,1821+643,
and PG\,0953+415, using high-resolution STIS data
(Richter et al.\,2004\,[9], Sembach et al.\,2004\,[7], Richter et al.\,2005,
in preparation) and have detected 26 candidate systems
along a total redshift path of $\Delta z=0.939$.
Fig.\,1 shows
an example of a broad Ly\,$\alpha$ absorber at
$z=0.18047$ in the STIS spectrum of H\,1821+643.
The detected broad Ly\,$\alpha$ systems have H\,{\sc i} column densities of
$12.7\leq $ log $N\leq 14.0$ and $b$ values ranging
from $40$ to $133$ km\,s$^{-1}$. 
Our measurements imply a number of broad Ly\,$\alpha$
systems per unit redshift of $dN_{\rm BL}/dz =28 \pm 5$
for $z<0.4$ and a cosmological mass
density of $\Omega_{b}$(BL)$\approx 0.0058\,h_{75}^{-1}$, which
is roughly twice the value obtained for the
O\,{\sc vi} absorbers.
For this preliminary estimate we have
omitted the contributions traced
by the broadest Ly\,$\alpha$ absorbers with $b>100$ km\,s$^{-1}$.
A large fraction of the baryons in the low redshift 
Universe may exist in those absorbers because the
implied temperatures and ionization corrections are
very large.  However, absorbers with $b>100$ km\,s$^{-1}$ are
difficult to study because of their weakness and large
width. Note that for the above estimate of $\Omega_{b}$(BL) we
also have not considered any
non-thermal line broadening mechanisms that
may contribute to the observed large $b$ values.
Such non-thermal processes include broadening by the
Hubble flow, peculiar gas motions, and macroscopic
turbulence. Future observational and theoretical 
investigations of $\Omega_{b}$(BL) will need to
address the influence of these processes for
the determination of $T$ from the measured line widths.
This will be crucial to test whether our current estimate
of the baryon content of the broad Ly\,$\alpha$ absorbers
is realistic.

\end{document}